\documentclass[aps,groupedaddress]{revtex4}

\usepackage{graphicx}
\usepackage{latexsym}
\usepackage{amssymb}
\usepackage{epsfig}

\begin{document}

\title{
Dosimetry for radiocolloid therapy of cystic craniopharyngiomas
}

\author{E. Leticia Rojas, Feras M.O. Al-Dweri and Antonio M. Lallena}

\affiliation{Departamento de F\'{\i}sica Moderna,
Universidad de Granada, E-18071 Granada, Spain.}

\author{Coral Bodineau and Pedro Gal\'an}

\affiliation{Servicio de Radiof\'{\i}sica Hospitalaria, Hospital
Regional Universitario ``Carlos Haya'', \\Avda. Carlos Haya s/n, E-29010
M\'alaga, Spain.}

\vspace*{1cm}

\begin{abstract}
The dosimetry for radiocolloid therapy of cystic craniopharyngiomas is
investigated. Analytical calculations based on the Loevinger and the
Berger formulae for electrons and photons, respectively, are compared
with Monte Carlo simulations. The role of the material of which
the colloid introduced inside the craniopharyngioma is made of as
well as that forming the cyst wall is analyzed. It is found that the
analytical approaches provide a very good description of the simulated
data in the conditions where they can be applied (i.e., in the case of a
uniform and infinite homogeneous medium). However, the consideration
of the different materials and interfaces produces a strong reduction
of the dose delivered to the cyst wall in relation to that predicted
by the Loevinger and the Berger formulae.
\end{abstract}


\maketitle

\section{Introduction}

Craniopharyngiomas are tumors showing an incidence of 3\% for all
tumors in adults and 6 to 10\% of tumors in children
\cite{Rub72}. Though histologically benign, they are effectively
malignant because they usually appear in a situation which can affect
to important organs such as hypothalamus, optic nerves or chiasms.

Treatment based on surgery followed (or not) by radiation therapy
needs a total resection to guarantee a good chance of cure. However, a
complete excision is rarely possible due to the vicinity of the
organs mentioned above \cite{Sha79}.

Alternatively, the introduction of radioactive colloids into the cyst
has been considered since the early 50's \cite{Lek53,Wyc54}. Cyst
wall constitutes the target volume and, as a consequence,
$\beta$-emitters, as e.g. $^{32}$P and $^{90}$Y are the most
adequate. Also $\beta \gamma$-emitters, as $^{186}$Re and $^{198}$Au, have
been used, since $\gamma$-radiation allows to test the
colloid distribution by obtaining a gamma camera image.

Cystic craniopharyngiomas are roughly spherical and the wall of the
cyst presents a thickness between 1 and 3 mm. Once the radiocolloid is
introduced inside the craniopharyngioma, part of it is attached to the
inner surface of the cyst, while the rest appears to be distributed into
the inner volume.

Dosimetry calculations for $\beta$ emitters have been carried out
either on the base of the Loevinger formulation \cite{Loe56} (see
e.g. \cite{NMT99}) or by using the $\beta$ point kernels generated by
Berger \cite{Ber71} (see e.g. \cite{Mcg86}). In the case of $\beta
\gamma$-emitters, the dose rates for photons have been evaluated
following the approach of Berger \cite{Ber68} in terms of the build-up
factors for the appropriate energies (see e.g. \cite{Mcg86}).

In order to perform the dosimetry,
because of the characteristic distribution of the radiocolloid, two
extreme situations have been considered in practice  \cite{NMT99,Mcg86}.
The first one is the spherical shell source, where the
emitter is supposed to be uniformly distributed in the inner surface
of the cyst. The second one
is the spherical volume source (SVS), which assumes the
radionuclide uniformly distributed in the inner volume of the
craniopharyngioma.  In all cases, a uniform and infinite homogeneous
medium is considered and the presence of different material media and
interfaces is not taken into account.

Monte Carlo (MC) simulation offers the possibility to perform a more
realistic dosimetry. In this paper we have studied how the
consideration of the actual media involved in the craniopharyngiomas
modifies the results of the standard calculations.

\section{Material and Methods}

Here we have assumed that the cystic craniopharyngioma is
described by means of a spherical layer of thickness $\delta$ and
inner radius $R$ (see Fig. 1). We adopted SVS approach and
then we dealt with extended uniform spherical sources of given
radionuclides. We were interested in calculating the dose delivered to
the cyst wall as well as to points external to the craniopharyngioma
and related to the critical surrounding tissue. The corresponding dose
rate is accounted for by evaluating the integral
\begin{equation}
\dot{D}(x) \, = \, \int_{\rm V} \, {\rm d} \textbf{r} \, C(\textbf{r}) \,
                   J( \textbf{x}-\textbf{r} ) \,
              = \, A \, \int_{\rm V} \, {\rm d}\textbf{r} \,
               J(\left| \textbf{x}-\textbf{r} \right|) \, ,
\label{eq:dose-rate}
\end{equation}
where V represents the source volume, $\textbf{x}$ and $\textbf{r}$
are, respectively, the positions of the target point and of the volume
element of the distribution with respect to its center,
$C(\textbf{r})\equiv A$ gives the radionuclide activity concentration
and $J(\left| \textbf{x}-\textbf{r} \right|) $ represents the absorbed
dose rate in $\textbf{x}$ due to a point source in $\textbf{r}$.

\begin{figure}
\begin{center}
\epsfig{file=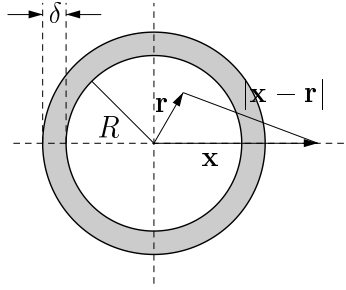,width=5cm}
\caption{\small 
Model scheme used to describe the craniopharyngioma.}
\end{center}
\end{figure}

\subsection{$\beta$ sources}

In the case of $\beta$ sources, dosimetry calculations can be performed
on the basis of the Loevinger formula \cite{Loe56}, which gives the absorbed
dose distribution around a point source of a $\beta$-emitting
radionuclide in a homogeneous infinite medium. In this approach the
absorbed dose rate at a distance $x$ from the point source is given by
\begin{equation}
J_\beta(x) \, = \, T_1(x) \, + \, T_2(x) \, ,
\label{eq:loev}
\end{equation}
where
\begin{equation}
T_1(x) \, = \, \left[ \frac{B t}{x^2} \, - \,
\frac{B}{x} \, \exp \left( 1-\frac{x}{t} \right) \right] \,
\Theta(t-x)
\label{eq:t1}
\end{equation}
and
\begin{equation}
T_2(x) \, = \, \frac{B}{x} \, \exp \left( 1-\frac{x}{z} \right) \, .
\label{eq:t2}
\end{equation}
In previous equations, $z=(\rho \nu)^{-1}$ and $t=cz$ are
characteristic distances, $\rho$ the density of the homogeneous
medium, $\nu$ an apparent absorption coefficient and $c$ a
dimensionless parameter. The normalization constant $B$ is
evaluated by imposing that all the emitted energy is absorbed in a
very large sphere, and is given by
\[
B \, = \, \displaystyle \frac{1}{4\pi}
 \, \rho \, \nu^2 \, \alpha \, \overline{E}_\beta \, ,
\]
with
\[
\alpha \, = \, \left[ 3c^2 \, - \, (c^2 - 1 ) \exp(1) \right]^{-1}
\]
and $\overline{E}_\beta$ the average energy per disintegration. Finally,
\[
\Theta(r) \, = \, \left\{
\begin{array}{ll}
0 \, ,  & r<0 \, ;\\
1 \, ,  & r>0 \, .
\end{array}
\right.
\]

The parameters $\nu$ and $c$ are characteristic of each $\beta$-emitter. For
the first one, Loevinger {\it et al.} \cite{Loe56} gave the following numerical
parameterization:
\[
\nu \, = \, \displaystyle
\frac{18.6}{\left( E_\beta^{\rm max} - 0.036 \right)^{1.37}} \,
\left( 2-\frac{\overline{E}_\beta}{\overline{E}^*_\beta} \right)
\, ,
\]
where $E_\beta^{\rm max}$ is the maximum energy of the electrons
emitted by the source, in MeV, and $\nu$ is in cm$^2$~g$^{-1}$.
$\overline{E}^*_\beta$ is the average energy for a hypothetical
$\beta$ disintegration of allowed type with the same maximum
energy. The $c$ parameter was parameterized as \cite{Loe56}:
\[
c \, = \, \left\{
\begin{array}{ll}
2   \, ,   & 0.17~{\rm MeV} < E_\beta^{\rm max} < 0.5~{\rm MeV} \, ;\\
1.5 \, ,   & 0.5~{\rm MeV} \leq E_\beta^{\rm max} < 1.5~{\rm MeV} \, ;\\
1   \, ,   & 1.5~{\rm MeV} \leq E_\beta^{\rm max} < 3~{\rm MeV}  \, .
\end{array}
\right.
\]

Adopting this approach for the point absorbed dose rate, the integral in
Eq. (\ref{eq:dose-rate}) can be evaluated analytically (see Appendix)
and using Eq. (\ref{eq:LL-fin}) we can write

\begin{eqnarray}
\label{eq:Loevfinal}
\dot{D}_\beta(x) & = & A \, \left\{
\displaystyle \frac{\pi B t}{2 x} \left( t \left[ 12 x - 11 t +
4 (t - R) \exp \left( \displaystyle \frac{t+R-x}{t} \right)
\right] \right. \right. \\
&& \left. ~~~~~~+ \, 3 R^2 + 2 R x - 5 x^2
+ 2 (R^2 - x^2) \ln \left( \displaystyle \frac{t}{x-R} \right)
\right) \, \Theta(R+t-x) \nonumber \\
&& \left. ~~~~+  \displaystyle \frac{2 \pi B z^2}{x} \exp
\left( \frac{z - R - x }{z} \right) \left[ (R+z)+ (R-z)
\exp \left( \frac{2 R}{z} \right) \right]  \right\} \, \Theta(x-R)
\nonumber \, .
\end{eqnarray}

\subsection{Photon sources}

When $\beta \gamma$-emitters are used, the photon dosimetry is usually
based on the formula of Berger \cite{Ber68}. In this approach, a
photon point isotropic source is assumed to deposit a dose rate at a
distance $x$ from the source which is given by
\begin{equation}
J_\gamma(x) \, = \, \frac{1}{4\pi} \,
\frac{\mu_{\rm en}}{\rho} \, E_\gamma
\, \frac{1}{x^2} \, \exp (-\mu x) \, B_{\rm en} (\mu x) \, .
\label{eq:ber}
\end{equation}
Here $\mu_{\rm en}$ and $\mu$ are the linear photon energy-absorption
and the linear photon attenuation coefficients, respectively, at
the energy of the emitted photon, $E_\gamma$, and $B_{\rm en}$ is the
energy-absorption buildup factor, which takes into account the
contribution of the scattered photons. Following Ref. \cite{Ber68},
the buildup factors are expanded as
\begin{equation}
B_{\rm en} (\mu x) \, = \, \sum_{n=0}^{10} \, b_n \, (\mu x)^n \, .
\label{eq:buildup}
\end{equation}
Here $b_0=1$ and the remaining coefficients were calculated in
\cite{Ber68} for some energies ranging from 10~keV to 3~MeV.

As in the case of the Loevinger formula, the integral in
Eq. (\ref{eq:dose-rate}) can be calculated analytically for the dose
rate in Eq. (\ref{eq:ber}), in the case of the uniform spherical
radionuclide distribution we are assuming here (see Appendix). From
Eq. (\ref{eq:F2}) we have
\begin{equation}
\dot{D}_\gamma(x) \, = \,
- A \, \frac{\mu_{\rm en}}{\rho} \, \frac{1}{4x}
\, E_\gamma \, \sum_{j=0}^{N+3} \, r_j(x) \, K_j(x) \, ,
\label{eq:Bergfinal}
\end{equation}
where the different terms are defined in Eqs. (\ref{eq:rrr})-(\ref{eq:ggg}).

\subsection{Simulation procedure}

MC simulations were performed by using the code PENELOPE \cite{Sal01}.
PENELOPE is a general purpose MC code which allows to simulate the
coupled electron-photon transport. Analog simulation is performed for
photons. For electrons, the simulation is carried out in a mixed
scheme where collisions are characterized as hard and soft. After
fixing a value for a critical angle, the collisions with scattering
angles larger than the critical value are called hard collisions and
are simulated individually. The collisions with scattering angle
smaller than the critical value are called soft collisions and are
described by means of a multiple scattering theory. The electron
tracking is controlled by four parameters. Two of them, called $C_1$
and $C_2$, refer to elastic collisions. The first one, $C_1$, gives the
average angular deflection due to an elastic hard collision and to the
soft collisions previous to it. The second parameter, $C_2$,
represents the maximum value permitted for the average fractional
energy loss in a step. The other two parameters, called $W_{\rm cc}$
and $W_{\rm cr}$ are energy cutoffs to distinguish hard and soft
events. Thus, the inelastic electron collisions with energy loss
$W<W_{\rm cc}$ and the emission of Bremsstrahlung photons with energy
$W<W_{\rm cr}$ are considered in the simulation as soft interactions.

For arbitrary materials,
PENELOPE can be applied for energies up to 1~GeV and down
to few hundred eV, in the case of electrons, and 1~keV, for photons.
Besides, PENELOPE permits a good description of
the particle transport at the interfaces and presents a more accurate
description of the electron transport at low energies in comparison to
other general purpose MC codes. These characteristics make PENELOPE to
be an useful tool for medical physics applications as previous works
have pointed out (see e.g. Refs. \cite{San98}-\cite{Ase02}). Details
about the physical processes considered can be found in
Ref. \cite{Sal01}.

In our simulations, the cyst was supposed to be surrounded by an
infinitely extended water volume. Both, the spherical layer simulating
the craniopharyngioma as well as the colloid introduced into the cyst
were assumed to be made of different materials.  Thus, we labelled the
dose rates obtained with the simulations as $\dot{D}_\omega^{\rm vs}$,
where the subscript $\omega$ refers to $\beta$ or $\gamma$ according to the
primary particle emitted and the superscripts v and s indicate,
respectively, the materials of which the inner volume and the
spherical shell are made of. Here we considered four materials: soft
tissue (t), compact bone (b), gel (g) and water (w). The respective
compositions are given in Table~\ref{tab:compos}.

\begin{table}[hb]
\begin{tabular}{cccc}
\hline \hline
& \multicolumn{3}{c}{Composition [atoms/molecule]} \\\hline
& Soft tissue (ICRU) & Compact bone (ICRU) & Gel \\\hline
Density [g~cm$^{-3}$] & 1.0 & 1.85 & 1.2914 \\ \hline
H  & 0.10117 & 0.52790 & 0.54697 \\
C  & 0.11100 & 0.19247 & 0.23524 \\
N  & 0.02600 & 0.01603 & 0.05393 \\
O  & 0.76183 & 0.21311 & 0.16156 \\
Mg & & 0.00068 & \\
P  & & 0.01879 & \\
S  & & 0.00052 & 0.00230 \\
Ca & & 0.03050 & \\
\hline \hline
\end{tabular}
\caption{Composition of the different materials used in the MC
simulations performed in this paper.
\label{tab:compos}}
\end{table}

The point sources representing the radionuclide were assumed to be
uniformly distributed inside the craniopharyngioma. The direction of
the emitted particles was supposed to be isotropic around the initial
position. For electrons, the initial energy was sampled from the
corresponding Fermi distributions, taken from Ref. \cite{Gar85}. For
photons, the initial energy was sampled according to the relative
intensity of the different energies of the corresponding spectra.

Electrons and photons were simulated for energies above 100~eV and
1~keV, respectively. Below these energies, the particles were considered
to be locally absorbed. The simulation parameters were fixed to the values:
$W_{\rm cc}=5$~keV, $W_{\rm cr}=1$~keV, $C_1=C_2=0.05$.

Due to the spherical symmetry of the adopted SVS geometry, the dose
distributions were supposed to be functions of the distance to the
center of the distribution, $x$. The full simulation volume was
subdivided into spherical shells with radial thickness of 0.5~mm. The
energy deposited in each voxel was scored to obtain the corresponding
histograms.

The statistical uncertainties were calculated by scoring both the
energy deposited in a voxel and its square for each history. The
average energy deposited in the $k$-th bin (per incident particle)
is
\[
 E_k \, = \, \frac{1}{N} \sum_{i=1}^{N} e_{i,k} \, ,
\]
where $N$ is the number of simulated histories and $e_{i,k}$ is the
energy deposited by all the particles of the $i$-th history (that is,
including the primary particle and all the secondaries it
generates). The statistical uncertainty is given by
\[
 \sigma_{E_k} \, = \,
 \sqrt{\frac{1}{N} \left[ \frac{1}{N}\sum_{i=1}^{N} e_{i,k}^2
\, - \, E_k^2 \right] } \, .
\]
In our calculations, $5 \cdot 10^6$ histories were simulated in
each run.

\section{Results}

\subsection{Pure $\beta$ radionuclides}

We started by considering a pure $\beta$ emitter as radionuclide in
the colloid: the $^{32}$P nucleus. The $\beta$-spectrum of this isotope has a
maximum energy $E_\beta^{\rm max}=1.71$~MeV with a mean energy
$\overline{E}_\beta=0.695$~MeV. The decay transition to $^{32}$S is of
allowed type and the parameters entering in the Loevinger formula are
$c=1$ and $\nu=9.18$~cm$^2$~g$^{-1}$. Initially we analyzed cysts with
$R=1.75$~cm and $\delta=1$~mm, while the activity concentration
of the radiocolloid was supposed to be 1~MBq~cm$^{-3}$.

\begin{figure}
\begin{center}
\epsfig{file=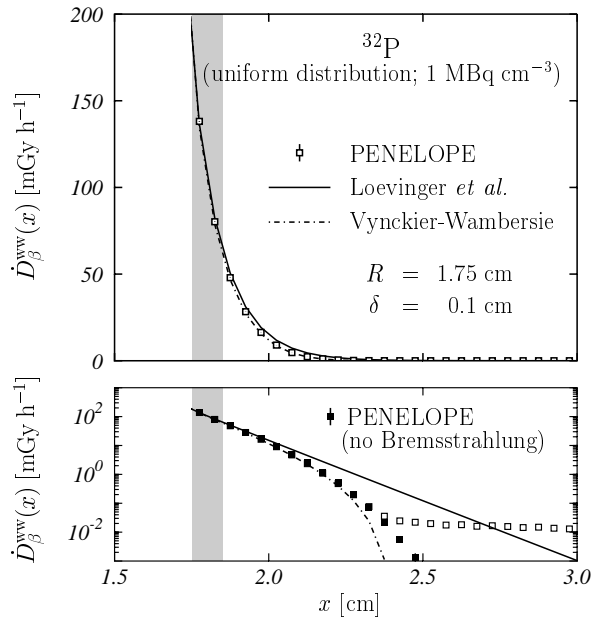,width=8cm}
\caption{\small Dose rate in water for a spherical source of $^{32}$P
with $R=1.75$~cm, $\delta=0.1$~cm and an activity concentration of
1~MBq~cm$^{-3}$. The distance of the target point to the center of the
distribution is labelled $x$. Dots are for the MC calculation (Error
bars correspond to 1$\sigma$ and are smaller than the size of the
symbol used). Solid curve gives the result obtained from the Loevinger
formula while dashed-dotted curve correspond to the Vynckier-Wambersie
formula. The lower panel shows the same results in semilogarithmic
scale. Besides the MC results obtained switching it off Bremsstrahlung
are plotted with black squares.}
\end{center}
\end{figure}

First, we have assumed that both the inner volume and the layer were
made of water. In this situation (uniform and infinite homogeneous
medium), the results of the simulation can be compared directly with
our analytical calculations based on the Loevinger approach
(Eq. (\ref{eq:Loevfinal})) or other analytical approaches such as that
of Vynckier and Wambersie \cite{Vyn82}. In this last method the
$\beta$ point kernels generated were more accurate than those in the
Loevinger approach. The results are shown in the upper panel of
Fig. 2, where the full and dashed curves show the results obtained
with the Loevinger and Vynckier and Wambersie approaches,
respectively, and the open squares are the results of our MC
simulation. In the lower panel, the same results are plotted in
semilogarithmic scale in order to emphasize the differences at large
distances. In Table \ref{tab:diff-ww} we show the values of the
relative differences calculated as:
\begin{equation}
\Delta^{\rm (A,B)}_\beta(x) \, = \, \displaystyle
\frac{ \left[ \dot{D}_\beta^{\rm ww}(x) \right]_{\rm A} \, - \,
\left[ \dot{D}_\beta^{\rm ww}(x) \right]_{\rm B} }
{\left[ \dot{D}_\beta^{\rm ww}(x) \right]_{\rm B} } \, ,
\label{eq:diff-ww}
\end{equation}
where the subscripts A and B stand for the Loevinger (L) and Vynckier
and Wambersie (VW) approaches or the MC simulation. As we can see,
Loevinger and MC results agree rather well up to $x \sim 2$~cm. Above
this point, the disagreement is apparent, but the value of the dose
rate is, at least, one order of magnitude smaller than the dose rate
in the cyst wall. Vynckier and Wambersie model provides a slightly
better result up to $x \sim 2.3$~cm. At larger distances, the Vynckier
and Wambersie approach goes to zero, but there the dose rate is, at
least, three orders of magnitude below that in the wall. Similar
results are obtained if the $\beta$ point kernels used are those
provided by Berger \cite{Ber71}.

\begin{table}[hb]
\begin{tabular}{cccc}
\hline \hline
& \multicolumn{3}{c}{$\Delta^{\rm (A,B)}_\beta(x)$ [\%]} \\
\hline
$x$ [cm]  & (L,MC) & (VW,MC) & (VW,L) \\
\hline
  1.775 & -1.1 & -2.8 & -1.7 \\
  1.825 & 2.6 & -2.0 & -4.4 \\
  1.875 & 5.5 & -2.6 & -7.7 \\
  1.925 & 10.0 & -3.8 & -12.5 \\
  1.975 & 17.2 & -5.3 & -19.2 \\
  2.025 & 30.7 & -5.7 & -27.8 \\
  2.125 & 92.4 & -5.1 & -50.6 \\
  2.225 & 258.3 & -17.2 & -76.9 \\
  2.325 & 754.5 & -68.3 & -96.3 \\
  2.425 & 921.5 & -100.0 & -100.0 \\
  2.525 & 386.8 & -100.0 & -100.0 \\
  3.025 & -93.6 & -100.0 & -100.0 \\
  3.525 & -99.9 & -100.0 & -100.0 \\
  4.025 & -100.0 & -100.0 & -100.0 \\
  4.525 & -100.0 & -100.0 & -100.0 \\
  5.025 & -100.0 & -100.0 & -100.0 \\
\hline \hline
\end{tabular}
\caption{Relative differences (see Eq. (\protect\ref{eq:diff-ww})) in
\% between the results obtained for $^{32}$P, in the case of unique
medium, with the Loevinger (L) and Vynckier and Wambersie (VW)
approaches and MC simulation.
\label{tab:diff-ww}}
\end{table}

To finish the discussion for a unique medium, we analyzed the role of
the Bremsstrahlung radiation. PENELOPE includes this mechanism and, in
order to estimate its importance we have performed a new simulation by
switching it off. The results obtained are plotted with black squares
in the lower panel of Fig. 2. As we can see, the effect of the
Bremsstrahlung shows up above $\sim 2.4$~cm. However, at that
distance, the dose deposited is four orders of magnitude smaller than
the dose delivered to the cyst wall. On the other hand, we can see that
the Vynckier and Wambersie model does not include Bremsstrahlung at
all, while the Loevinger approach takes it into account in a somehow
average way. Despite this, the conclusion is that these analytical
approaches permit a reasonable description of the dose rate delivered
by this kind of extended sources within a unique medium.

The next step was to investigate the role played by the interfaces
present in the craniopharyngioma. The idea was to elucidate, by using
MC simulation, if the various interfaces and materials modify the
dosimetry provided by the Loevinger approach. To measure the effect of
the presence of the different materials, we calculated the relative
differences
\begin{equation}
\Delta^{\rm v's'-vs}_{\omega}({\rm cyst}) \, = \, \displaystyle
\frac{\dot{D}_\omega^{\rm v's'}({\rm cyst}) \, - \,
\dot{D}_\omega^{\rm vs}({\rm cyst})}
{\dot{D}_\omega^{\rm vs}({\rm cyst})} \, ,
\label{eq:reldif}
\end{equation}
where 
\begin{equation}
\dot{D}_\omega^{\rm vs}({\rm cyst}) \, = \,
\displaystyle \frac{1}{\rho_{\rm s} v_{\rm s}} \sum_{k} E_k 
\label{eq:integ-dose}
\end{equation}
are the integral dose rates delivered inside the cyst wall.
Here $k$ runs over all the bins contained inside the cyst wall,
$\rho_{\rm s}$ is the density of the material forming this wall and
$v_{\rm s}$ is its volume.

\begin{table}[hb]
\begin{tabular}{cccccccc}
\hline \hline
&& \multicolumn{6}{c}{$\Delta^{\rm v's'-vs}_\beta({\rm cyst})$ [\%]} \\
\hline
$R$ [cm]  & $\delta$ [mm] &
wt-ww & wb-ww & gt-wt & gb-wb & gt-ww & gb-ww  \\
\hline
1.75 & 1.0 & 0.87  & -28.02 & -20.67 & -20.61 & -19.99 & -42.86 \\
1.75 & 2.0 & 1.18  & -39.65 & -21.15 & -21.24 & -20.22 & -52.47 \\
1.75 & 3.0 & 0.98  & -44.77 & -20.81 & -20.99 & -20.03 & -56.37 \\
1.00 & 1.0 & 0.46  & -27.39 & -20.62 & -20.58 & -20.25 & -42.34 \\
2.50 & 1.0 & 0.50  & -28.44 & -20.50 & -21.02 & -20.11 & -43.48 \\
\hline \hline
\end{tabular}
\caption{Relative differences in \%, as given by
Eq. (\protect\ref{eq:reldif}), for $^{32}$P in the different cases
analyzed in this paper (see text).
\label{tab:reldif}}
\end{table}

We have first studied the effect of the material forming the spherical
shell which represents the craniopharyngioma.  We have considered two
extreme cases: soft tissue and compact bone. The differences of the
dose rates calculated with MC in both situations with respect to
$\dot{D}_\beta^{\rm ww}(x)$ are shown in panels {\it (a)} and {\it
(b)} of Fig. 3. The relative differences in the cyst wall
are given in Table \ref{tab:reldif}. As we can see, the presence of the
soft tissue in the spherical shell leaves practically unmodified the
dosimetry. In fact, the total dose rate delivered to this shell
increases by only 0.9\% in this case. On the contrary, when the bone
is considered  the dose received by the wall reduces 
28.0\%. Furthermore, the dose rate delivered to the external medium
nearby the craniopharyngioma diminishes appreciably in the case of the bone,
while no modifications are observed for soft tissue.  Obviously, these
modifications are not considered in the analytical expressions based
on the Loevinger approach.

\begin{figure}
\begin{center}
\epsfig{file=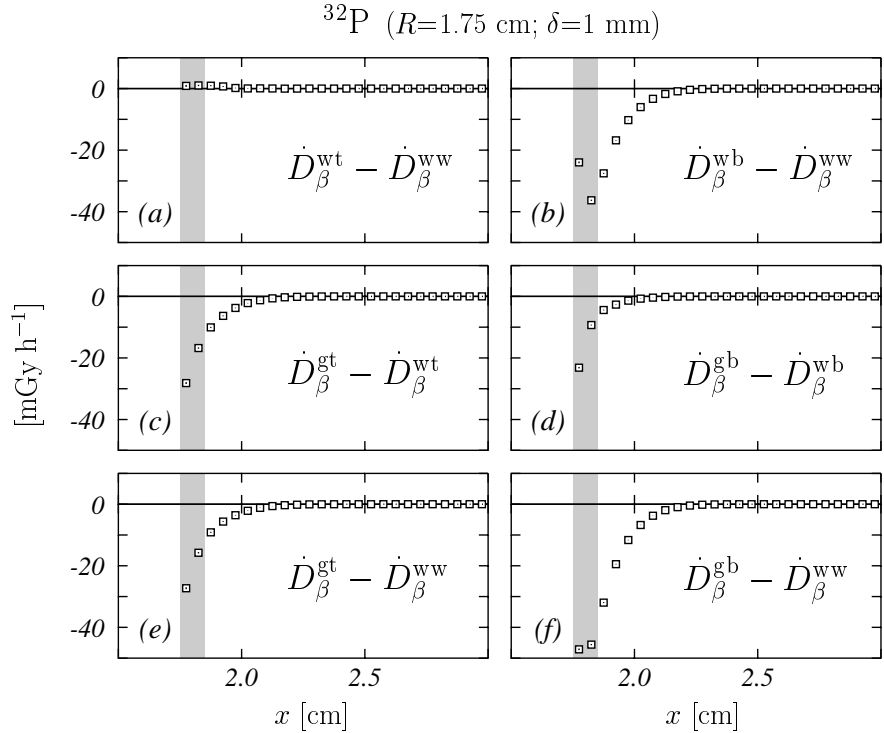,width=12cm}
\caption{\small Effect of the materials considered for the spherical
shell which represents the cystic wall in our model and for the inner
volume. Panels {\it (a)}\/ and {\it (b)}\/ show the effect of
considering the wall of the craniopharyngioma to be done of soft
tissue and bone, respectively. Panels {\it (c)}\/ and {\it (d)}\/ show
the effect due to the substitution of the water by the gel in the
inner volume. Finally, panels {\it (e)}\/ and {\it (f)}\/ show the
full effect due to the presence of these materials with respect to the
reference dose rate $\dot{D}_\beta^{\rm ww}(x)$. The shadow region
shows the cystic wall which has been considered to have a thickness
$\delta=1$~mm. Error bars correspond to 1$\sigma$ and are smaller than
the size of the symbol used.}
\end{center}
\end{figure}

In the treatment procedure, the radionuclide is introduced in the
craniopharyngioma inner volume, embedded in a certain type of colloid.
In our simulations we have considered the gel with the composition
given in Table~\ref{tab:compos}. In panels {\it (c)}\/ and {\it (d)}\/
of Fig. 3 we show the differences in the dose rates
obtained when the water is substituted by the gel in the inner
volume. As we can see, the gel produces a noticeable reduction of the
dose delivered to the cyst wall. This reduction reaches 20.7\% in
the case of soft tissue and 20.6\% in the case of the bone. Besides, also the
dose delivered to the surrounding outer volume reduces.

The net effect due to the presence of both gel in the inner volume and
soft tissue or bone in the craniopharyngioma, in relation to the case
in which only water is considered, is shown in panels {\it (e)}\/ and
{\it (f)}\/ of Fig. 3. The dose rate delivered to the cyst
wall reduces in 20.0\%, in the case of the soft tissue, and 42.9\%
in the case of bone. Even if we can expect that the craniopharyngioma is
composed of a material with an intermediate density, the dose rate
delivered to both, the cyst wall and the outer volume nearby it,
would be strongly reduced in relation to what it is obtained using the
Loevinger approach.

\begin{figure}
\begin{center}
\epsfig{file=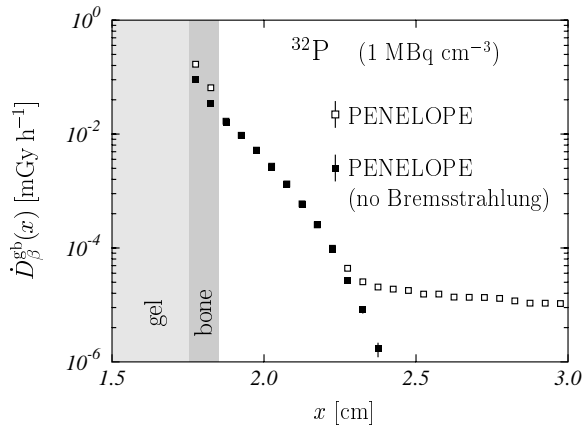,width=8cm}
\caption{\small Effect of the Bremsstrahlung radiation for the case
in which gel and bone are present in the inner volume and the wall of
the craniopharyngioma, respectively. Black squares give the dose rate
obtained when Bremsstrahlung mechanism is switched off. Open squares
correspond to the full calculation.}
\end{center}
\end{figure}

The presence of materials with densities larger than that of water,
could make the Bremsstrahlung effect more important. As before, we
have carried out a new simulation (for the case in which gel and bone
are present in the inner volume and in the wall of the craniopharyngioma,
respectively), but we switched off the Bremsstrahlung mechanism. In
Fig. 4 the results obtained (black squares) are compared to those
of the full calculation (open squares). These results
are rather similar to those found for the
case of a unique medium (see lower panel in Fig. 2), except for two
details. First, the Bremsstrahlung effect begins to be observed in the
radiation queue at a slightly shorter distance, $\sim 2.2$~cm. Second,
the inclusion of Bremsstrahlung effects produces an increase of the dose
delivered to the cyst wall which reaches the non negligible value of $\sim
45$\%. If we compare the simulation without Bremsstrahlung with the
analytical approaches which, as mentioned previously, do not treat
this effect in an adequate way, the reduction observed for
the dose delivered to the cyst wall would be even larger.

At this point, it is necessary to clarify how the previous
conclusions are related to the choice of the geometry. For this reason
we
have performed a set of calculations similar to the previous ones but
assuming different thicknesses for the tumor wall and different sizes
for the inner volume.

First we have analyzed the effect of the wall thickness and we have
considered $\delta=2$ and 3~mm. The relative differences $\Delta^{\rm
v's'-vs}_\beta({\rm cyst})$ obtained are shown in Table
\ref{tab:reldif} for different materials and configurations. As we can
see, the effect of the presence of the soft tissue is rather small,
increasing the dose delivered to the cyst by $\sim 1$\% (see third
column). On the contrary, the fact that the wall is made of bone
produces a considerable reduction of the dose rate in it. Besides,
this reduction grows with the thickness of the cyst wall (see fourth column).

The effect of the presence of the gel in the inner volume seems to be
almost independent of the thickness of the wall producing a reduction of
$\sim 20$\% in all cases (see fifth and sixth columns).

Finally, the combined effect of soft tissue in the wall and of the gel in the
inner volume (seventh column) is practically equal to that of the gel alone,
while in the case of bone (eight column), the reduction increases with
the wall thickness, following the behavior observed when no gel was
considered. It is worth to point out that the dose rate delivered to the
cyst can be diminished by more than 50\%.

To study the influence of the inner volume of the craniopharyngioma we
have considered two different values of its radius: $R=1$ and
2.5~cm. In both calculations we have maintained the activity
concentration of 1~MBq~cm$^{-3}$.  The results obtained in the
simulations are shown in the two last rows of Table
\ref{tab:reldif}. As we can see, the values of the relative
differences are very similar to those found for the former calculation
($R=1.75$~cm, $\delta=1$~mm). This indicates that the important
parameter is the thickness of the cyst wall. The
reduction of the dose delivered to it becomes bigger with the increase
of the thickness.

\subsection{$\beta \gamma$ sources}

After having analyzed the case of pure $\beta$ sources, we studied
$\beta \gamma$ sources. To perform the calculations we have considered
the $^{186}$Re radionuclide which has the decay scheme shown in Fig. 5
\cite{Chu99}. Following Ref. \cite{Ann98}, in order to simplify the
simulation, we considered only the most relevant transitions, whose
properties are given in Table \ref{tab:renio}.  In this table, we show
for X-rays, internal conversion (IC) and Auger electrons (these last
two have been grouped) the average values of the energies and the
intensities. The two $\beta$ decays considered are first forbidden
(not unique). However, the sampling procedure \cite{Gar85} is the same
as for allowed transitions.

\begin{figure}
\begin{center}
\epsfig{file=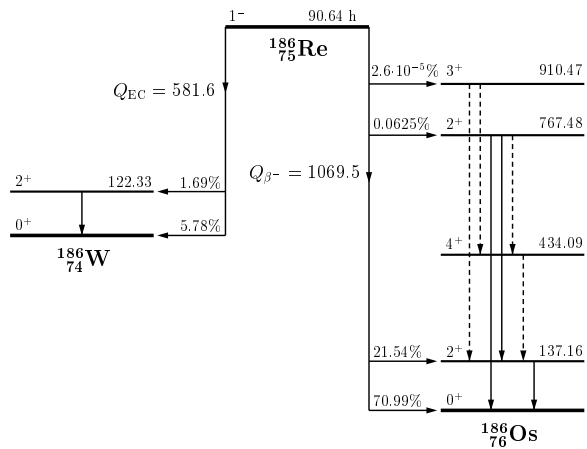,width=8cm}
\caption{\small Decay scheme of the $^{186}$Re radionuclide. In actual
simulations, only the $\beta$ decays populating the ground and the
first excited state of the $^{186}$Os, the $\gamma$ decays shown with
solid lines as well as the X-rays and IC and Auger electrons mentioned
in the text have been considered.}
\end{center}
\end{figure}

\begin{table}[hb]
\begin{tabular}{ccc}
\hline \hline
Type  & Energy [keV] & Absolute intensity [\%] \\
\hline
$\beta^-$ & 1069.50 & 70.99 \\
$\beta^-$ & ~932.34 & 21.54 \\
\hline
$\gamma$  & ~767.50 & ~0.03 \\
$\gamma$  & ~630.34 & ~0.03 \\
$\gamma$  & ~137.16 & ~9.42 \\
$\gamma$  & ~122.30 & ~0.60 \\
\hline
W K X-rays & ~$\sim$65.00 & ~6.00 \\
Os K X-rays & ~$\sim$65.00 & ~3.50 \\
\hline
IC+Auger & ~$\sim$14.00 & ~22.00 \\
\hline \hline
\end{tabular}
\caption{Characteristic radiations considered in this paper for the
decay of the $^{186}$Re. The energies given for the $\beta^-$ are the
endpoint energies. Data have been obtained from
Refs. \protect\cite{Chu99,Ann98}.
\label{tab:renio}}
\end{table}

As in the previous case, we assumed a source of $R=1.75$~cm with
$\delta=1$~mm and an activity concentration of 1~MBq~cm$^{-3}$
introduced in the inner volume of the craniopharyngioma.

We began by comparing to the theoretical predictions based on the
Loevinger and Berger approaches, our MC simulations, done by assuming
a unique medium (water).

For the $\beta$ decay with $E_\beta^{\rm max}=1.069$~MeV, the mean
energy is $\overline{E}_\beta=0.362$~MeV and the parameters entering
in the Loevinger formula are $c=1.5$ and
$\nu=17.78$~cm$^2$~g$^{-1}$. For that with $E_\beta^{\rm
max}=0.932$~MeV, we used $\overline{E}_\beta=0.309$~MeV, $c=1.5$ and
$\nu=21.61$~cm$^2$~g$^{-1}$. The results corresponding to
$\dot{D}_\beta^{\rm ww}(x)$ are shown in Fig. 6.  The calculations
have been done by using the absolute intensities given in Table
\ref{tab:renio}. As we can see, the simulation (dots) and the
analytical calculation (solid lines) show a good agreement, as it
happened in the case of $^{32}$P.

\begin{figure}
\begin{center}
\epsfig{file=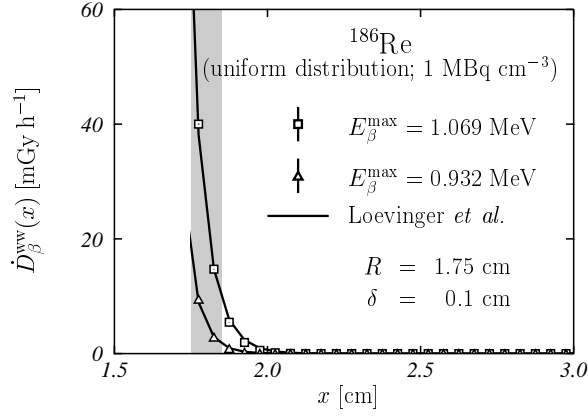,width=8cm}
\caption{\small Same as Fig. 2 but for the two main $\beta$ decays of
$^{186}$Re. Only the Loevinger approach results are shown. The
absolute intensities of both emissions are included.}
\end{center}
\end{figure}

\begin{figure}
\begin{center}
\epsfig{file=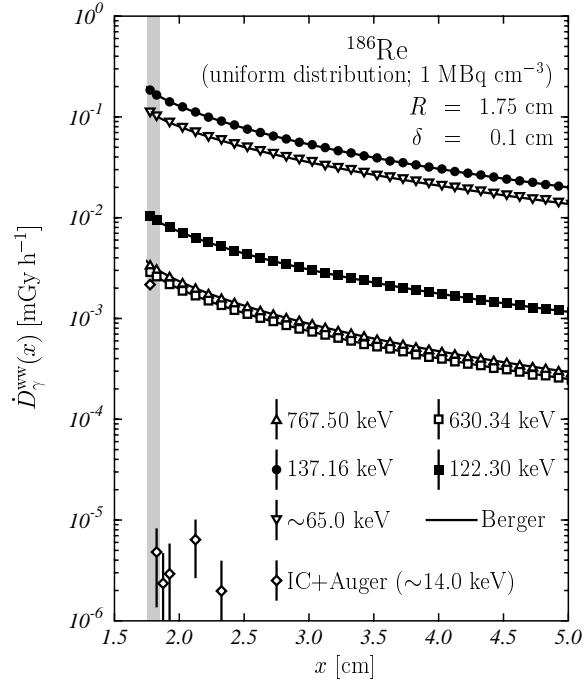,width=8cm}
\caption{\small  Same as Fig. 2 but for the main $\gamma$ and X-ray
radiations of $^{186}$Re. Solid curves show the result obtained from
the Berger approach. The absolute intensities of the different
emissions are included.}
\end{center}
\end{figure}

The results obtained for the remaining radiations in Table
\ref{tab:renio} are shown in Fig. 7. Also in these results,
the absolute intensities of
each radiation are included. For photons (these are the four $\gamma$
emissions and the X-rays), the results of the simulations (dots) are
compared with those provided by the analytical approach of Berger
(solid lines). The parameters entering in Eq. (\ref{eq:F2}),
and given in Table \ref{tab:param}, have been
obtained by performing cubic interpolations of the values quoted in
Ref. \cite{Ber68}. It is
remarkable the excellent agreement between the analytical approach and
the MC simulation. It is also worth to notice that the $\gamma$
emissions with $E_\gamma=137.16$~keV and the X-ray of W and Os are the
most important contributions here, while the other emissions are more than
one order of magnitude smaller.

\begin{table}[htb]
\begin{tabular}{cccccc}
\hline \hline
$E_\gamma$ [keV] & 65 & 122.30 & 137.16
 & 630.34 & 767.50 \\ \hline
 $\mu$ [cm] & 0.191  & 0.158   & 0.153   & 0.087  & 0.080 \\
 $\mu_{\rm en}$ [cm]
            & 0.0292 & 0.02613 & 0.02691 & 0.0328 & 0.0322 \\
\hline
 $b_1$    & ~2.50964594E+00 & ~1.77904308E+00 & ~1.6780888E+00
& ~9.16518331E-01 & ~8.59057903E-01 \\
 $b_2$    & ~1.42232013E+00 & ~1.10754991E+00 & ~9.9470758E-01
& ~3.82239223E-01 & ~3.56937379E-01 \\
 $b_3$    & ~1.86807606E-02 & ~1.91051394E-01 & ~1.8541029E-01
& ~4.75710258E-02 & -1.91028565E-02 \\
 $b_4$    & ~1.71705126E-03 & -5.35412366E-03 & -6.1609102E-03
& -4.75984300E-03 & -4.22946505E-05 \\
 $b_5$    & -9.58392775E-06 & ~6.26657449E-04 & ~7.3442928E-04
& ~9.06498346E-04 & ~3.30697017E-04 \\
 $b_6$    & ~3.01566342E-06 & -1.99327806E-05 & -3.3124285E-05
& -8.57197738E-05 & -4.32139241E-05 \\
 $b_7$    & -2.95050882E-07 & ~1.93571609E-07 & ~1.0333482E-06
& ~4.61301079E-06 & ~2.68527947E-06 \\
 $b_8$    & ~1.26480826E-08 & ~1.07387210E-08 & -1.9661881E-08
& -1.42838076E-07 & -9.03993893E-08 \\
 $b_9$    & -2.53243121E-10 & -3.77658183E-10 & ~1.9686815E-10
& ~2.36108977E-09 & ~1.57900648E-09 \\
$b_{10}$  & ~1.96279971E-12 & ~3.65828331E-12 & -7.2484147E-13
& -1.60724663E-11 & -1.11767913E-11 \\
\hline \hline
\end{tabular}
\caption{Parameters to be used in the analytical approach based on the
Berger formula discussed in this paper. The medium considered is
water. The different radiations correspond to the $\gamma$ and X-ray
emissions taken into account for $^{186}$Re. The value of the
parameter $b_0$ is fixed to 1 for all energies.
\label{tab:param}}
\end{table}

\begin{table}[htb]
\begin{tabular}{cccc}
\hline \hline
& & \multicolumn{2}{c}{$\Delta^{\rm v's'-vs}_{\beta,\gamma}({\rm cyst})$ [\%]} \\
\hline
Type  & Energy [keV] & gw-ww & gb-ww \\ \hline
$\beta^-$ & 1069.50     & -20.37 & -52.21 \\
$\beta^-$ & ~932.34     & -20.90 & -54.34 \\
$\gamma$  & ~767.50     & ~-0.33 & ~-3.26 \\
$\gamma$  & ~630.34     & ~-1.86 & ~-4.30  \\
$\gamma$  & ~137.16     & ~-0.97 & ~31.55 \\
$\gamma$  & ~122.30     & ~-1.51 & ~47.30 \\
X-rays    & $\sim 65.0$ & ~-0.24 & 217.38 \\
IC+Auger  & $\sim 14.0$ & -20.25 & -57.89 \\ \hline
\multicolumn{2}{c}{Full spectrum} & -20.26 & -51.19 \\
\hline \hline
\end{tabular}
\caption{Relative differences in \%, as given by
Eq. (\protect\ref{eq:reldif}), for the different radiations of
$^{186}$Re. Here $R=1.75$~cm and $\delta=1$~mm. Last row gives the
results for the full spectrum of $^{186}$Re considered in this
paper. The energies given for the $\beta^-$ are the endpoint energies.
\label{tab:reldif-Re}}
\end{table}

In Fig. 7 we also show the results the IC and Auger
electrons (diamonds). These phenomena cannot be described by
the analytical approaches considered here but, in any case,
their contribution seems to be negligible.

\begin{figure}
\begin{center}
\epsfig{file=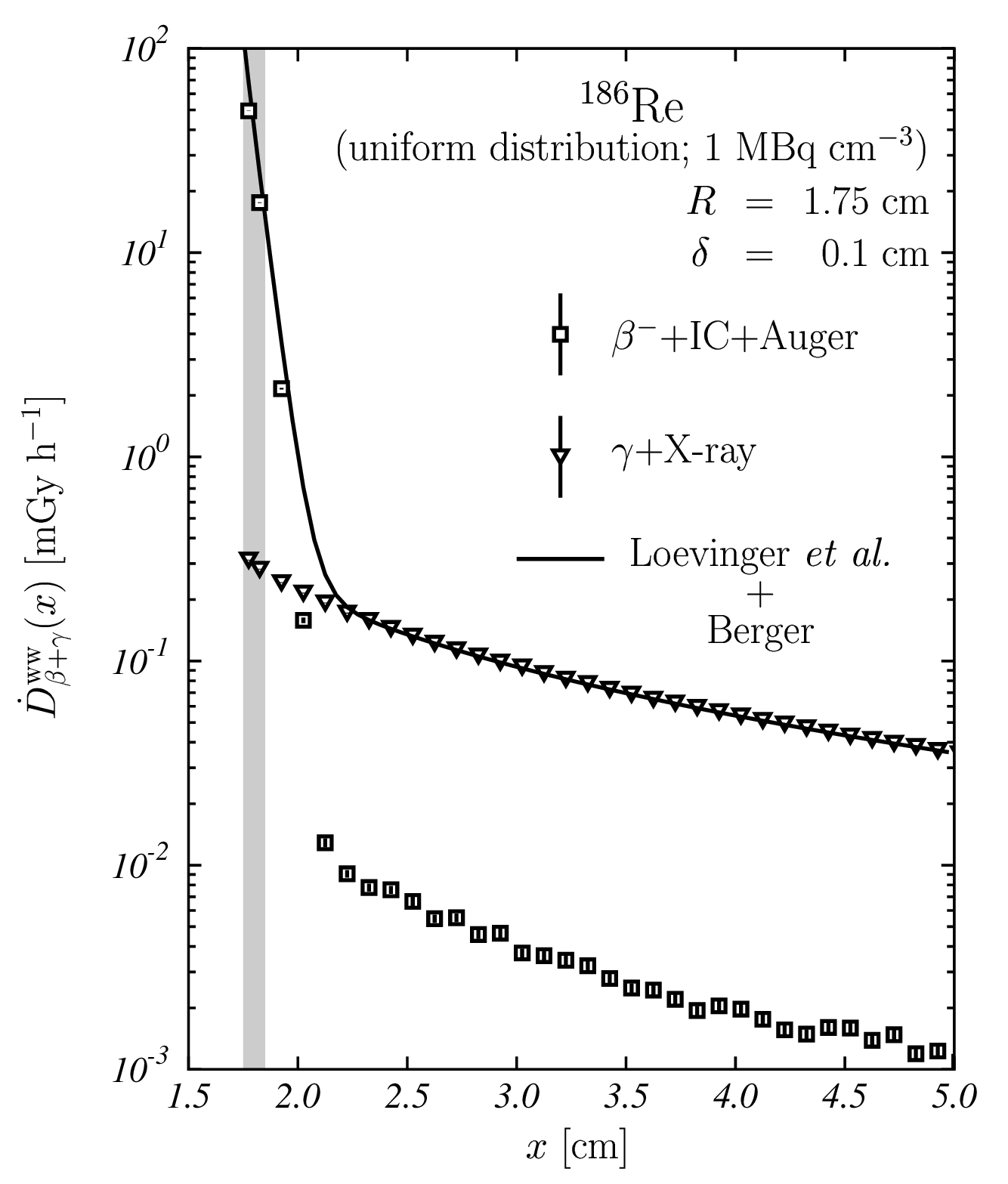,width=8cm}
\caption{\small Same as Fig. 2 but for the full spectrum emission of
$^{186}$Re. The separate contribution of the $\beta$ emission plus IC
and Auger electrons (squares) and that of $\gamma$ plus X-rays
(triangles) are shown. Solid curve shows the analytical result
obtained from the Loevinger and Berger approaches.}
\end{center}
\end{figure}

In Fig. 8 we show the separate contributions of the electrons,
say $\beta$ emissions, IC and Auger electrons (squares), and of the
photons, $\gamma$ emissions and X-rays (triangles). The solid curve
corresponds to the analytical calculation performed by summing the
contributions obtained with the Loevinger approach for the two $\beta$
emissions and with the Berger approach for the $\gamma$ emissions and
the X-rays. As we can see it provides a rather good description of the
simulated results. Besides, it is worth to point out that the dose
delivered to the craniopharyngioma is mainly due to the $\beta$
radiation, the photon contribution being two orders of magnitude
smaller. On the contrary, the dose due to electrons becomes more than
one order of magnitude smaller than the one delivered by $\gamma$
emissions and X-rays for $x>2$~cm, that is at distances above 2~mm far
from the outside surface of the craniopharyngioma.

Finally, we have tested the effect of the different materials and
interfaces in the dosimetry. First, we have evaluated the dose rate by
assuming gel in the inner volume of the craniopharyngioma. The results
corresponding to the relative differences, as given by
Eq. (\ref{eq:reldif}), are those in the third column of Table
\ref{tab:reldif-Re}. In this table we also show the values for each of the
radiations considered in the spectrum of the $^{186}$Re and, in the
last row, for the full spectrum (by including the proper absolute
intensities). As we can see, the presence of the gel produces a
reduction of the dose delivered to the cyst wall. This reduction is of
$\sim 20$\% for the $\beta$ emissions and for the IC and Auger
electrons. The reduction, no greater than 2\%, is very small for
$\gamma$ and X-ray radiations. The full effect of the gel is
20.3\% reduction, as expected due to the dominance of the $\beta$
radiations previously discussed.

In a second step we have studied the effect of the material forming
the cyst wall. We have checked that, as it happened for the $^{32}$P
radionuclide, the presence of soft tissue, instead of water, produces
negligible modifications. In the last column of Table
\ref{tab:reldif-Re} we show the results obtained when compact bone is
assumed to constitute the cyst wall. As we can see, the reduction of
the dose delivered to the cyst reaches more than 50\% for the $\beta$,
IC and Auger electron emissions. However, the behavior for $\gamma$
radiations is different. While for the more energetic photons the dose
reduces (less than 5\%), it increases considerably for the less
energetic ones, with an enormous enhancement (more than 200\%) for
X-rays. In any case, the dominance of the $\beta$ emissions
on the full spectrum is again remarkable.

The average energies of the $\beta$ emissions in this radionuclide are
smaller than these of $^{32}$P. This implies that Bremsstrahlung effect
is less important in this case. On the other hand, we have check that
the Bremsstrahlung effect for the $\gamma$ emissions is negligible,
the dose rates obtained without Bremsstrahlung overlapping with those
found in the full calculation.

To finish, we compare the integral dose rates  (see
Eq. (\ref{eq:integ-dose})) delivered by the two
radionuclides here considered  inside the cyst wall.
In the case of a unique medium (water) and
assuming the same radionuclide concentration injected in the inner
volume of the cyst, the dose rate deposited in the wall by the
$^{198}$Re is 33\% smaller than the one due to $^{32}$P. When we
consider the cyst filled with gel and the cyst shell made of bone,
the dose rate of $^{198}$Re is 25\% smaller than that of $^{32}$P.

\section{Conclusions}

In this paper we have investigated the dosimetry for radiocolloid
therapy of cystic craniopharyngiomas. The craniopharyngioma has been
described as a spherical shell with given internal radius and
thickness. Different materials have been considered to form the
colloid introduced in the craniopharyngioma and the shell representing
the cyst wall. Explicit analytical expressions for the dose rate due
to $\beta$ and $\gamma$ emitters have been obtained on the base of the
well known Loevinger and Berger formulae for point sources. Results for
$^{32}$P and $^{186}$Re are quoted.

The results obtained with these analytical expressions,
valid under the assumption of a unique medium,
have been compared
with calculations performed with the Monte Carlo simulation code PENELOPE.
These calculations
show a nice agreement for both $\beta$ and $\gamma$ emissions in
the two radionuclides studied.

By using the Monte Carlo simulation, the role of the different
materials and of the interfaces present in the problem, has been
investigated. The main conclusions we draw are the following ones.
\begin{enumerate}

\item The effect of the material forming the colloid (gel in our case)
produces a reduction of $\sim 20$\% of
the dose delivered to the cyst wall by
$\beta$ and IC and Auger electron emissions.
This reduction is smaller than 2\% for $\gamma$ and X-ray
emissions.

\item There are no noticeable differences between the results obtained
when soft tissue or water are considered to form the cyst wall.

\item The consideration of compact bone as the material forming the
cyst wall gives rise to relevant effects. For $\beta$ and IC and Auger
electron emissions, the dose delivered to the cyst wall reduces by a factor
larger than 25\%. This effect together with that produced by the
colloid makes this dose to be reduced by more than 40\%. For
$\gamma$ and X-ray emissions, the presence of the bone in the cyst
wall appears to be very important for energies below 200~keV. In this
situation, the dose delivered to the cyst wall increases
by more than 200\% in the case of the X-rays of 65~keV we have considered.
For the $^{186}$Re the dominant emissions are
$\beta$ radiations, and they obscure this enormous effect.

\item All the effects quoted above, do not change with the size (inner
radius) of the craniopharyngioma, but increase with its thickness,
which appear to be the relevant parameter to consider in the
dosimetry.

\end{enumerate}

\appendix
\section{Formulae for spherical uniform sources}

In this appendix we discuss the procedure to calculate analytically
the integral in Eq. (\ref{eq:dose-rate}) for the point dose rates
corresponding to the Loevinger and Berger approaches.

In the first case, and taking into account Eq. (\ref{eq:loev}), the
dose rate per activity unit at a distance $x$ from the center of the
distribution can be written as
\begin{equation}
\frac{1}{A} \,  \dot{D}_\beta(x) \, = \, I_1(x) \, + \, I_2(x) \, ,
\label{eq:ext3}
\end{equation}
where
\begin{equation}
I_i(x) \, = \, \int_{\rm V} \, {\rm d}\textbf{r} \,
T_i(\left| \textbf{x}-\textbf{r} \right|) \, .
\end{equation}

\begin{figure}[hb]
\begin{center}
\epsfig{file=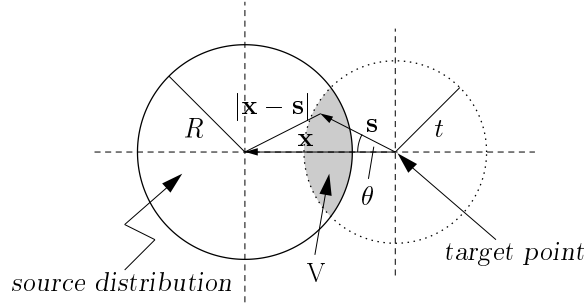,width=8cm}
\caption{\small Geometrical scheme to calculate the integrals of the
Loevinger and Berger formulae for an extended spherical source uniformly
distributed.}
\end{center}
\end{figure}

The key point to evaluate the corresponding integrals is to fix the
center of the coordinate system on the target point. The scheme is
represented in Fig. 9. In the case of the integral $I_1(x)$, one has
to consider that $T_1(x)$ has a range $t$ (see
Eq. (\ref{eq:t1})). Thus, the only contribution to the integral comes
from points inside the source distribution and sited at a distance $s
\leq t$ from the target point. Therefore, only the shadowed volume
shown in Fig. 9 contributes to $I_1(x)$ and we have
\begin{eqnarray}
\nonumber
I_1(x) & = & \int_{\rm V} \, {\rm d}\textbf{s} \, T_1(s) \, = \,
\int_{x-R}^t \, {\rm d}s \, s^2 \, T_1(s) \,
\int_0^{\theta_{\rm max}} \, {\rm d}\theta \sin \theta \,
\int_0^{2\pi} \, d\phi \\ &=&
\nonumber
2 \pi \, \int_{x-R}^t \, {\rm d}s \, s^2 \, T_1(s) \,
\int_0^{\cos \theta_{\rm max}} \, {\rm d}(-\cos \theta) \\ &=&
2 \pi \, \int_{x-R}^t \, {\rm d}s \, s^2 \, 
\left( 1-\cos \theta_{\rm max} \right) \, T_1(s) \, .
\label{eq:i1-uno}
\end{eqnarray}
The angle $\theta_{\rm max}$ is the value reached by $\theta$ when
$|\textbf{x}-\textbf{s}|=R$ and then
\begin{equation}
\cos \theta_{\rm max} \, = \, \displaystyle
\frac{x^2 + s^2 - R^2}{2 x s} \, .
\end{equation}
Considering the above equations, the integral (\ref{eq:i1-uno}) becomes
\begin{equation}
I_1(x) \, = \, \displaystyle -\frac{\pi}{x} \,
\Theta(x-R) \, \Theta(R+t-x) \,
\int_{x-R}^{t} {\rm d}s \, s \,
\left[ (x-s)^2 - R^2 \right] \, T_1(s)
\, .
\label{int1}
\end{equation}
Here, the $\Theta(x-R)$ factor indicates that the calculation is valid
for points outside the distribution, while the $\Theta(R+t-x)$ factor
points out the fact that this contribution cancels for $x$ values
larger than $R+t$. The result of the integration is
\begin{eqnarray}
\nonumber
I_1(x) & = & \displaystyle \frac{\pi B t}{2 x} \left\{
t \, \left[ 12 x - 11 t +
4 (t - R) \exp \left( \displaystyle \frac{t+R-x}{t} \right)
\right] \, + \, 3 R^2 + 2 R x - 5 x^2
\right. \\ && \left.
 ~~~~~~~~+ 2 (R^2 - x^2) \ln \left( \displaystyle
\frac{t}{x-R} \right)
\right\} \, \Theta(x-R) \, \Theta(R+t-x) \, .
\label{eq:I1final}
\end{eqnarray}
Note that, in the limit $x=R$ this contribution is
\begin{equation}
I_1(R) \, = \, \displaystyle \frac{\pi B t^2}{2R}
\left\{ 4 R \left[3-\exp(1) \right]
 - t \left[ 11-4 \exp(1) \right] \right\} \, .
\end{equation}

To calculate the contribution $I_2(x)$, a similar procedure is
followed. Now, the full source distribution contributes and this
implies that in Eq. (\ref{int1}) we must change the upper
limit in the integral to $x+R$. Thus one has
\begin{eqnarray}
I_2(x) & = & \displaystyle -\frac{\pi}{x} \,
\int_{x-R}^{x+R} {\rm d}s \, s \,
\left[ (x-s)^2 - R^2 \right] \, T_2(s) \nonumber \\
& = &
\displaystyle \frac{2 \pi B z^2}{x} \exp
\left( \frac{z - R - x }{z} \right)
 \left[ (R+z)+ (R-z)
\exp \left( \frac{2 R}{z} \right) \right]
\, .
\label{eq:I2final}
\end{eqnarray}

By substituting Eqs. (\ref{eq:I1final}) and (\ref{eq:I2final}) into
Eq. (\ref{eq:ext3}) we have finally
\begin{eqnarray}
\label{eq:LL-fin}
\frac{1}{A} \,  \dot{D}_\beta(x) & = & \left\{
\displaystyle \frac{\pi B t}{2 x} \left(
t \left[ 12 x - 11 t +
4 (t - R) \exp \left( \displaystyle \frac{t+R-x}{t} \right)
\right] \right. \right. \\
&& \nonumber \left.
~~~~~~~~~~+ \, 3 R^2 + 2 R x - 5 x^2
+ 2 (R^2 - x^2) \ln \left( \displaystyle \frac{t}{x-R} \right)
\right) \, \Theta(R+t-x) \\
&& \nonumber
 \left. +  \displaystyle \frac{2 \pi B z^2}{x} \exp
\left( \frac{z - R - x }{z} \right)
\left[ (R+z)+ (R-z)
\exp \left( \frac{2 R}{z} \right) \right]  \right\} \, \Theta(x-R) \, .
\end{eqnarray}

For $\gamma$ emitters, the integral in Eq. (\ref{eq:dose-rate}) is
evaluated for the point dose rate given by the Berger approach
(\ref{eq:ber}). In this case, $J_\gamma(x)$ has no range and the
calculation follows the same steps as those for the $T_2$ term of the
Loevinger formula. Thus one can write
\begin{equation}
\frac{1}{A} \,  \dot{D}_\gamma(x) \, = \,
\int_{\rm V} \, {\rm d}\textbf{r} \,
J_\gamma(\left| \textbf{x}-\textbf{r} \right|) \, = \,
\displaystyle -\frac{\pi}{x} \,
\int_{x-R}^{x+R} {\rm d}s \, s \,
\left[ (x-s)^2 - R^2 \right] \, J_\gamma(s).
\end{equation}
Substituting Eqs. (\ref{eq:ber}) and (\ref{eq:buildup}) here one has
\begin{equation}
\frac{1}{A} \,  \dot{D}_\gamma(x) \, = \,
-\frac{\mu_{\rm en}}{\rho} \, \frac{1}{4x}
\, E_\gamma \, \sum_{j=0}^{N+3} \, r_j(x) \, K_j(x) \, ,
\label{eq:F2}
\end{equation}
where
\begin{equation}
r_j(x) \, = \,
\sum_{i={\rm max}(0,j-N-1)}^{{\rm min}(j,2)} \,
a_i(x) \, b_{j-i} \, \mu^{j-i} \, ,
\label{eq:rrr}
\end{equation}
with
\begin{equation}
a_i(x) \, = \, \left\{
\begin{array}{ll}
x^2 \, - \, R^2 \, , & i=0 \, , \\
-2x \, , & i=1 \, , \\
1 \, , & i=2 \, ,
\end{array}
\right.
\end{equation}
and
\begin{equation}
K_j(x) \, = \, \int_{x-R}^{x+R} \, {\rm d}s \,
s^{j-1} \, \exp (-\mu s) \, ,
\end{equation}
If $j=0$ the integral $K_0(x)$ is given by \cite{Abr72}
\begin{eqnarray}
K_0(x) &=& E_1 \left( \mu (x-R) \right) \, - \,
    E_1 \left( \mu (x+R) \right) \\
\nonumber
&=& \ln \displaystyle \frac{x+R}{x-R} \, + \,
\sum_{n=1}^{\infty} \, \frac{(-1)^n \, \mu^n}{n \, n!}
\, \left[ (x+R)^n \, - \, (x-R)^n \right] \, ,
\end{eqnarray}
where $E_1(x)$ is the exponential integral \cite{Abr72}.
For $j \not = 0$ we have \cite{Abr72}
\begin{equation}
K_j(x) \, = \, \displaystyle \frac{1}{\mu^j} \,
\left[ \gamma \left( j,\mu(x+R) \right) \, - \,
       \gamma \left( j,\mu(x-R) \right) \right] \, ,
\,\,\, j\not = 0 \, ,
\end{equation}
with
\begin{equation}
\gamma (n,z) \, = \, (n-1)! \, P(n,z) \, = \, (n-1)! \, \left\{
1 \, - \, \exp (-z) \, \sum_{k=0}^{n-1} \,
\frac{z^k}{k!} \, .
\right\}
\label{eq:ggg}
\end{equation}

\acknowledgments{ 
We acknowledge G. Co' for the careful reading of the manuscript.
E.L.R. acknowledges the financial support of the
I.N.I.N. (Mexico). F.M.O. A.-D. acknowledges the A.E.C.I. (Spain) and
the University of Granada for funding his research stay in Granada
(Spain).  This research has been supported in part by the Junta de
Andaluc\'{\i}a (FQM0225).
}


\begin{thebibliography}{99}

\bibitem{Rub72} L.J. Rubinstein, {\it Tumors on the central nervous
system} (Armed Forces Institute of Pathology, Washington DC, 1972).

\bibitem{Sha79}
K. Shapiro, K. Till and N. Grant, ``Craniopharyngiomas in childhood,''
J. Neurosurg. 50, 617-623 (1979).

\bibitem{Lek53} L. Leksell and L.A. Liden, ``A therapeutic trial with
radioactive isotopes in cystic brain tumor,'' in {\it Radioisotope
Techniques. Vol. 1: Medical and Physiological Applications}
(Her Majesty's Stationery Office, London, 1953) p. 76.

\bibitem{Wyc54}
H.T. Wycis, R. Robbins, M. Spiegel-Adolph, J. Meszaros and E.A. Spiegel,
``Treatment of a cystic craniopharyngioma by injection of radioactive P-32,''
Confin. Neurol. 14, 193-202 (1954).

\bibitem{Loe56}
R. Loevinger, E.M. Japha and G.L. Brownell, ``Discrete radioisotope sources,''
in {\it Radiation Dosimetry}, edited by G.J. Hine and G.L. Brownell
(Academic Press, New York, 1956) pp. 694-799.

\bibitem{NMT99}
J.C. Harbert, J.S. Robertson and K.D. Held,
{\it Nuclear medicine therapy} (Thieme Medical Publishers, New York, 1987).

\bibitem{Ber71}
M.J. Berger, ``Distribution of absorbed dose around point sources of
electrons and beta particles in water and other media,''
MIRD Pamphlet No. 7, J. Nucl. Med. Suppl. 5, 7-23 (1971).

\bibitem{Mcg86}
E.L. McGuire, S. Balachandran and C.M. Boyd, ``Radiation dosimetry considerations
in the treatment of cystic suprasellar neoplasms,''
Br. J. Radiol. 59, 779-785 (1986).

\bibitem{Ber68}
M.J. Berger, ``Energy deposition in water by photons from point isotropic
sources,''  MIRD Pamphlet No. 2, J. Nucl. Med. Suppl. 1, 17-25 (1968).

\bibitem{Abr72}
M. Abramowitz and I.A. Stegun, {\it Handbook of mathematical functions}
(Dover Publ. Inc., New York, 1972).

\bibitem{Sal01}
F. Salvat, J.M. Fern\'andez-Varea, E. Acosta and J. Sempau, {\it PENELOPE, a
code system for Monte Carlo simulation of electron and photon
transport} (NEA-OECD, Paris, 2001).

\bibitem{San98}
A. S\'anchez-Reyes, J.J. Tello, B. Guix and F. Salvat, ``Monte Carlo
calculation of the dose distributions of two $^{106}$Ru eye
applicators,'' Radiother. Oncol. 49, 191-196 (1998).

\bibitem{Sem01}
J. Sempau, A. S\'anchez-Reyes, F. Salvat, H. Oulad ben Tahar, S.B. Jiang
and J.M. Fern\'andez-Varea, ``Monte Carlo simulation of electron beams from an
accelerator head using PENELOPE,'' Phys. Med. Biol. 46, 1163-1186 (2001).

\bibitem{Ase02}
J. Asenjo, J.M. Fern\'andez-Varea and A. S\'anchez-Reyes, ``Characterization of a
high-dose-rate $^{90}$Sr-$^{90}$Y source for intravascular
brachytherapy by using the Monte Carlo code PENELOPE,'' Phys. Med.
Biol. 47, 697-711 (2002).

\bibitem{ICRU}
International Commision on Radiation Units and Measurements,
ICRU Report No. 46, {\it Photon, electron, proton and neutron interaction
data for body tissues} (ICRU, Bethesda, 1992).

\bibitem{Gar85}
E. Garc\'{\i}a-Tora\~no and A. Grau Malonda, ``EFFY, a new program to
compute the counting efficiency of beta particles in liquid
scintillators,'' Comput. Phys. Commun. 36, 307-312 (1985).

\bibitem{Vyn82}
S. Vynckier and A. Wambersie, ``Dosimetry of beta sources in radiotherapy
I. The beta point source dose function,'' Phys. Med. Biol. 27, 1339-1347 (1982).

\bibitem{Chu99}
S.Y.F. Chu, L.P. Ekström and R.B. Firestone,
``WWW Table of Radioactive Isotopes, database version 2/28/99,''
from URL http://nucleardata.nuclear.lu.se/nucleardata/toi/

\bibitem{Ann98}
M.F. L'Annunziata (ed.), {\it Handbook of radioactivity analysis}
(Academic Press, San Diego, 1998).

\end{thebibliography}
\end{document}